\documentclass[fleqn]{annalen}
\usepackage{graphics}
\begin{document}
\newcommand{\volume}{11}              
\newcommand{\xyear}{2000}            
\newcommand{\issue}{5}               
\newcommand{\recdate}{15 November 1999}  
\newcommand{\revdate}{dd.mm.yyyy}    
\newcommand{\revnum}{0}              
\newcommand{\accdate}{dd.mm.yyyy}    
\newcommand{\coeditor}{ue}           
\newcommand{\firstpage}{507}         
\newcommand{\lastpage}{510}          
\setcounter{page}{\firstpage}        
\newcommand{\keywords}{Schwarzschild black hole, maximal slicings} 
\newcommand{\PACS}{04.20Cv} 
\newcommand{\shorttitle}
{R.\ Beig, Maximal slicing} 
\title{The maximal slicing of a Schwarzschild black hole}
\author{R.\ Beig} 
\newcommand{\address}
  {Institut f\"ur Theoretische Physik, Universit\"at Wien, Boltzmanngasse 5,
 A-1090 Wien, Austria  }
\newcommand{\email}{\tt beig@galileo.thp.univie.ac.at} 
\maketitle
\begin{abstract}
   We describe recent work due to Niall \'O Murchadha and the author 
(gr-qc/9706046, Phys. Rev. {\bf D57}, 4728 (1998)) on the late time
behaviour of the maximal foliation of the extended Schwarzschild geometry
which results from evolving a time symmetric slice into the past and future,
with time running equally fast at both spatial ends of the manifold.
We study the lapse function of this foliation in the limit where proper
time-at-infinity goes to ${+\infty}$ or ${-\infty}$ and the slices
approach $r=3m/2$.
\end{abstract}

\vspace*{0.25cm} \noindent
Talk given at the Journ\'ees Relativistes 1999, Weimar, Germany

\vspace*{0.5cm} \noindent
In this talk I report on work done jointly with Niall \'O Murchadha
\cite{BoM} where we rigoroulsy construct a foliation of the
$r > 3m/2$-region of the Kruskal spacetime by spherically symmetric,
asymptotically flat Cauchy surfaces with zero mean curvature. This slicing
is not new. It has previously been studied by Reinhart \cite{R},
Estabrook et al. \cite{E}, Smarr and York \cite{S}, Eardley and Smarr
\cite{Ea}, Brill et al. \cite{Br} and Petrich et al. \cite{P}.
We improve on these works by providing rigorous proof where these
authors use numerical or semi-analytical arguments.
In the process we also clarify some conceptual issues connected with the
different choices for slicing portions of the Kruskal spacetime by spherical
surfaces of zero mean curvature. The Kruskal spacetimes, in the regions
where either $r > 2m$ or $r < 2m$, can be written as
\begin{equation}
ds^2 = - \left(1 - \frac{2m}{r}\right) dt^2 +
\left( 1 - \frac{2m}{r}\right)^{-1} dr^2 + r^2 d\Omega^2,
\end{equation}
where $-\infty < t < \infty$ and $d\Omega^2$ is the line element of the unit
2-sphere. We seek spherically symmetric maximal slices which, locally, we
write as $\sigma =$~const where
\begin{equation}
\sigma = t - F(r).
\end{equation}
Note that spacelike spherical slices in the exterior region $r > 2m$ have
to be of this form whereas, for $r < 2m$, they could contain open sets
where $r$ is constant. But an easy computation shows that the mean
curvature of $r =$~const for $r < 2m$ is given by
\begin{equation}
K = \frac{1}{r^2 \sqrt{2m/r - 1}} (3m - 2r),
\end{equation}
which is only maximal when $r = 3m/2$. In fact, the slicing we are 
constructing will approach $r = 3m/2$ asymptotically.

The condition that $\sigma =$~const be maximal means that
\begin{equation}
K = \nabla_\mu n^\mu = 0,
\end{equation}
where
\begin{equation}
n_\mu dx^\mu = [-\nabla \sigma)^2]^{-1/2} (-dt + F'dr).
\end{equation}
Note that $N = [-(\nabla \sigma)^2]^{-1/2}$ is also given by
\begin{equation}
N = - n_\mu \xi^\mu ,
\end{equation}
where $\xi = \partial/\partial t$ is the static Killing vector. Thus $N$
can also be viewed as the ``boost function'' of $\xi$ relative to
$\sigma =$~const. Expressing $\frac{\partial}{\partial r} F = F'$ in
terms of $N$ we see that
\begin{equation}
F'{}^2 = \left( 1 - \frac{2m}{r}\right)^{-2} - 
\left( 1 - \frac{2m}{r}\right)^{-1} N^{-2}.
\end{equation}

Inserting (7) into Equ. (4) there follows
\begin{equation}
K = \frac{1}{r^2} \frac{\partial}{\partial r} 
\left[ r^2 \left(N^2 - \left( 1 - \frac{2m}{r}\right)\right)^{1/2}\right].
\end{equation}
Thus $N$ is of the form
\begin{equation}
N = \left( 1 - \frac{2m}{r} + \frac{C^2}{r^4}\right)^{1/2},
\end{equation}
where $C =$~const. Consequently
\begin{equation}
F'(r) = - \frac{C}{(1-2m/r)(r^4 - 2mr^3 + C^2)^{1/2}}.
\end{equation}
Consider, now, slices given by $\sigma = t - F(r,C) =$~const, where
\begin{equation}
F(r,C) = - \int_{r_C}^r \frac{C}{(1-2m/x)(x^4-2mx^3+C^2)^{1/2}}dx
\end{equation}
with $P(r_C) = r^4_C - 2m r^3_C + C^2 = 0$. Here we choose
$0 < C^2 < 27m^4/16$ and $3m/2 < r_C < 2m$. Note that $r_C \to 3m/2$
as $C$ approaches the critical value given by $3 \sqrt{3} m^2/4$.
Furthermore the integral in Equ.~(11) has to be understood in the
principal-value sense at $x = 2m$, so that $F$ has a logarithmic
singularity at $r = 2m$. In fact the (``Eddington--Finkelstein'') 
form of this singularity is exactly
that required in order for $\sigma =$~const to go smoothly through
$r = 2m$. Consider the slice given by $\sigma = 0$, which has the
property of being symmetric w.r. to the timelike cylinder given by
$t = 0$, which contains the sphere $r = r_C$.
Thus this slice goes to spacelike infinity at both spatial ends
of the Kruskal manifold, and time runs equally fast at these ends.
(The asymptotic values for $t$ at both ends differ by the amount
$2\tau(C)$, with $\tau(C)$ given by Equ. (12).)
Our slice can be varied in two ways. Firstly it can be
pushed along $\xi$, which corresponds to embedding it in the family
$\sigma =$ const: this does not yield a foliation since $\xi$
is tangential to $\sigma = 0$ at $r = r_C$. Secondly the parameter
$C$ can be varied: this is the choice we make. One can show \cite{BoM} 
that, for $C > 0$ this yields a foliation of the future half of the
Kruskal spacetime for $r > 3m/2$. Letting $C$ vary over 
$- 3 \sqrt{3} m^2/4 < C < 3 \sqrt{3} m^2/4$, one obtains a foliation
of Kruskal for $r > 3m/2$ (containing $t=0$ for $C=0$: we do not
investigate the smoothness of this slicing near $C = 0$, $r = 2m$). In
fact, rather
than $C$, we would like to take as parameter the quantity $\tau$ which is
proper time-at-infinity, namely
\begin{equation}
\tau(C) = - \int_{r_C}^\infty \frac{C}{(1 - 2m/x)(x^4 - 2mx^3 + C^2)^{1/2}}
dx.
\end{equation}
Note that $\tau \to \infty$ as $C \to 3 \sqrt{3}m^2/4$.
Now we compute the lapse function $\alpha$ of this foliation. The
unit-normal of this foliation, again denoted by $n^\mu$, is
\begin{equation}
n^\mu = - \alpha \nabla^\mu \tau.
\end{equation}
Writing $\xi$ as
\begin{equation}
\xi^\mu = N n^\mu + X^\mu , \qquad X^\mu n_\mu = 0 ,
\end{equation}
there follows, using Equ. (6), that
\begin{equation}
\xi^\mu \nabla_\mu \tau = N \alpha \alpha^{-2},
\end{equation}
whence
\begin{equation}
\alpha = N \frac{dt}{d\tau} = N \frac{dC}{d\tau} 
\frac{\partial F}{\partial C}.
\end{equation}
Thus we have to differentiate $F$ w.r. to $C$. This is not a trivial task.
Generalizing slightly a result due to Chow and Wang \cite{Ch}, we obtain
\begin{eqnarray}
\frac{\partial}{\partial C} F(r,C) &=&
\frac{1}{2(r-3m/2)(1-2m/r+C^2/r^4)^{1/2}} \nonumber \\
&& \mbox{} -
\frac{1}{2} \int_{r_C}^r \frac{x(x-3m)}{(x-3m/2)^2(x^4-2mx^3+C^2)^{1/2}}
dx,
\end{eqnarray}
which diverges at $r = r_C$. However $N \; \partial F/\partial C$ is 
regular at $r = r_C$. Namely we find that
\begin{eqnarray}
\alpha &=& \left( \frac{d\tau}{dC}\right)^{-1} 
\frac{1}{2} \left[ \frac{1}{r-3m/2} -
\left(1 - \frac{2m}{r} + \frac{C^2}{r^4}\right)^{1/2} \right.
\times \nonumber \\
&& \times \int_{r_C}^r \left. \frac{x(x-3m)}
{(x-3m/2)^2(x^4-2mx^3+C^2)^{1/2}} dx \right].
\end{eqnarray}
Note that $N$ and $\alpha$ are just linearly independent, spherical 
solutions of
\begin{equation}
(D^2 - K_{ij} K^{ij}) f = 0,
\end{equation}
where $N$ is antisymmetric and $\alpha$ is symmetric relative to the
``throat'' at $r = r_C$. We want to estimate the ``central'' lapse for
late times, i.e. $\alpha(r_C)$ for large $\tau$. From (18)
\begin{equation}
\alpha(r_C) = \left( \frac{d\tau}{dC}\right)^{-1} \frac{1}{2\delta},
\end{equation}
where $\delta = r_C - 3m/2$. Introducing the dimensionless variables
\begin{equation}
\bar \delta = \frac{\delta}{m}, \qquad
\bar \tau = \frac{\tau}{m}, \qquad
\bar C = \frac{C}{m^2},
\end{equation}
there holds
\begin{equation}
\bar \tau(\bar \delta) = - \bar C
\int_{3/2 + \bar \delta}^\infty \frac{y}{(y-2)(y^4-2y^3+\bar C^2)^{1/2}}
dy,
\end{equation}
and
\begin{equation}
\bar C = \left(\bar \delta + \frac{3}{2}\right)^{3/2}
\left(\frac{1}{2} - \bar \delta \right)^{1/2}.
\end{equation}
Letting $\bar \delta \to 0$, the polynomial $y^4 - 2y^3 + \bar C^2$
develops a double root at the lower limit where $y = 3/2$. By some 
tedious estimates \cite{BoM} we find
\begin{equation}
\bar \tau(\bar \delta) = - \frac{3\sqrt{6}}{4} \ln \bar \delta +
A + O(\bar \delta),
\end{equation}
where
\begin{equation}
A = \frac{3 \sqrt{6}}{4} \ln |18(3 \sqrt{2} - 4)| - 
2 \ln \left| \frac{3 \sqrt{3} - 5}{9 \sqrt{6} - 22}\right| = - 0,2181
\end{equation}
and
\begin{equation}
\frac{d \bar \tau}{d \bar C} = \frac{3}{4 \sqrt{2}} \frac{1}{\bar \delta^2}
+ O \left(\frac{1}{\bar \delta}\right).
\end{equation}
Consequently
\begin{equation}
\alpha(r_C) = \frac{4}{3 \sqrt{2}} \exp \frac{4A}{3\sqrt{6}}
\exp \left( - \frac{4\tau}{3 \sqrt{6} m} \right) 
 + O \left( \exp - \frac{8\tau}{3 \sqrt{6}m} \right)
\mbox{ as } \tau \to \infty
\end{equation}
which is our main result.

\vspace*{0.25cm} \baselineskip=10pt{\small \noindent I thank Dieter Brill
 for pointing out to me some references related to the topic of this work.
Work supported by Fonds zur F\"orderung der wissenschaftlichen Forschung,
Proj. Nr. P12626-PHY. }

\end{document}